\documentstyle[12pt]{article}

\input{psfig}

\newcommand{\lsim}{\mbox{ \raisebox{-1.0ex}{$\stackrel{\textstyle <}
{\textstyle \sim}$ }}}

\begin{document}

\vspace*{-10mm}

\baselineskip12pt
\begin{flushright}
\begin{tabular}{l}
{\bf KEK-TH-386 }\\
{\bf KEK preprint 93--204 }\\
{\bf CHIBA-EP-74 }\\
February 1994
\end{tabular}
\end{flushright}

\baselineskip18pt
\vspace{8mm}
\begin{center}
{\Large \bf Neutral Scalar Higgs Masses and \\
            Production Cross Sections in\\
            an Extended Supersymmetric Standard Model } \\
\vglue 8mm
{\bf Jun-ichi Kamoshita$^{1,2}$,
     Yasuhiro Okada$^1$
     and Minoru Tanaka$^1$}\\
\vglue 4mm
{\it Theory Group, KEK, Tsukuba, Ibaraki 305, Japan$^1$ }\\
{and} \\
{\it Department of Physics, Chiba University, Chiba, 263 Japan$^2$}\
\vglue 20mm
{\bf ABSTRACT} \\
\vglue 10mm
\begin{minipage}{14cm}
\baselineskip12pt
Upper bounds on the three neutral scalar Higgs masses are considered
 in the supersymmetric standard model with a gauge singlet Higgs field.
 When the lightest Higgs is singlet-dominated
 the second lightest Higgs is shown
 to lie near or below the theoretical upper bound on
 the lightest Higgs mass.
 We also consider detectability of these Higgs bosons
 at a future $e^+ e^-$ linear collider
 with $\sqrt{s}\sim 300 $ GeV and show
 that at least one of the neutral scalar Higgs
 has a production cross section
 larger than 0.04~pb.
\end{minipage}
\end{center}
\vfill
\newpage

In the standard model of the elementary particle physics
one of the most important subjects
both experimentally and theoretically is
 to explore the Higgs sector.
 In addition to establishing the symmetry breaking mechanism,
 an investigation of the Higgs sector gives us a clue
 to physics beyond the standard model.
 This might be the case for the supersymmetric (SUSY) standard model
 whose Higgs sector has distinct features.
In particular, a relatively light neutral Higgs
 is one of the general consequences of the SUSY models.
 Since no definite upper bound exists
 for the masses of the superpartners,
 the Higgs sector could be
 the first check point of the SUSY models.

In the minimal supersymmetric standard model (MSSM)
it is well-known that we can calculate
 the upper bound on the lightest neutral Higgs
 mass as a function of the top and the stop masses.
The upper bound is about 120 GeV for a top mass of 150 GeV
 and a stop mass of 1 TeV, and therefore
 exceeds the search limit of the LEP II experiment\cite{OYY}.

 If we consider extended versions of the SUSY standard model,
 the situation is a little different.
 The simplest and interesting extension is to include
 a gauge singlet Higgs field.
 In this case we can introduce a new coupling
 among two Higgs doublets
 and a singlet in the superpotential
 which induces a new quartic term
 in the tree level Higgs potential.
 Therefore, the tree level mass relations themselves
 are modified and there is no upper bound
 on the lightest neutral Higgs mass
 without specifying the strength
 of the new coupling constant.
 However, if we require
 that none of dimensionless coupling
 constants blow up below the grand unification scale,
 then we can determine
 an upper bound on the lightest neutral Higgs mass\cite{Drees,Ellis}.
 Taking account of the top and stop loop effects
 the upper bound is obtained as 130 $\sim$ 150 GeV
 for a reasonable range of the top mass and
 the 1 TeV stop mass\cite{singlet,Elliot}.

Although we can set a fairly strong mass bound
 for the lightest neutral Higgs,
 it may not be produced since it can be singlet-dominated
 in some region of parameter space
 and has a reduced coupling to the $Z^0$ boson
\cite{Elliot}.
 In such a case the bounds
 for the heavier Higgs masses become important.

In this letter we derive upper bounds on
the heavier neutral Higgs masses and show that
 the second lightest one obeys a similar
 but somewhat weaker mass bound
 if the lightest Higgs is dominated by a singlet component.
We also discuss the detectability of the neutral Higgses
 in a future $e^+ e^-$ linear collider like JLC-I\cite{JLC}
 and show that at least one of the Higgses has a production
 cross section for $e^+ e^- \rightarrow Z^0 h$ mode
 larger than 0.04~pb at $\sqrt{s}=300$ GeV.

We consider here
 a SUSY standard model with a gauge singlet Higgs field.
In order to obtain a general consequence of this model
 we introduce all possible terms allowed by the gauge symmetry
 and renormalizability in the superpotential
 and the SUSY soft breaking terms.
 The superpotential is given by\cite{Drees}
\begin{equation}
W=\lambda_1 H_1 H_2 N + \frac{1}{6}\lambda_2 N^3+\mu_1 H_1 H_2 +
  \frac{1}{2} \mu_2 N^2 ,
\end{equation}
then the tree level scalar potential is given by
\begin{eqnarray}
V&=&m_1^2|h_1|^2+m_2^2|h_2|^2+m_N^2|n|^2 \nonumber\\
 & &+[\lambda_1 A_1 nh_1h_2
    +\frac{1}{6}\lambda_2 A_1 n^3 +B_1 \mu_1 h_1h_2
     +\frac{1}{2}B_2 \mu_2 n^2 + h.c.] \nonumber \\
 & &+ (|h_1|^2+|h_2|^2)|\lambda_1 n
    + \mu_1|^2+|\lambda_1 h_1h_2+
      \frac{1}{2}\lambda_2 n^2+\mu_2 n|^2 \nonumber \\
 & &+ \frac{g^2}{8}(h_1^{\dagger}\tau^a h_1
    +h_2^{\dagger}\tau^a h_2)^2
    + \frac{g'^2}{8}(|h_1|^2-|h_2|^2)^2 ,
\end{eqnarray}
where, $H_1,H_2$ are doublet superfields
 and $N$ is a gauge singlet one,
and $h_1, h_2$ and $n$ are used
 for the scalar component\footnote{ In
 eq.(1) we have eliminated the linear term of $N$
 by a field redefinition of $N$.
 In the scalar potential we can in principle include
 a linear SUSY soft breaking term.
 But this does not make any difference
 in the following discussion.}.
 We have to include the top and stop loop effect
 to the above potential.
 For the degenerate stop case the one loop potential
 is given by
\begin {eqnarray}
V^{(1~loop)} & = &
 \frac{3}{16\pi^2}\Bigl[(m^2+y_t^2|h_2|^2)^2
(\ln{\frac{m^2+y_t^2|h_2|^2}{\mu^2}}-\frac{1}{2})\nonumber\\
 & & -y_t^4|h_2|^4
( \ln{\frac{y_t^2|h_2|^2}{\mu^2}}-\frac{1}{2})\Bigl] ,
\end {eqnarray}
where $m^2$ is a SUSY breaking mass for the squark,
 $\mu$ is a renormalization scale
 and $y_t$ is the top Yukawa coupling constant.
 In the case with the large left-right stop mixing
 the above formula should be extended appropriately
\footnote{We can also add the radiative correction due
to the $\lambda_1$ and $\lambda_2$ coupling constants but the effect is
not so large\cite{Elliot}.}.

{}From this potential we can calculate the Higgs mass matrix.
 In the present case physical states are three neutral scalars,
 two pseudoscalar and one charged Higgs pair.
 It is convenient to eliminate $m_1^2, m_2^2, m_N^2$
 by the minimization conditions of the potential
 and  use three vacuum expectation values,
 i.e. $v_1=\langle h_1^0\rangle,v_2=\langle h_2^0\rangle,x=\langle n\rangle$
 as independent parameters.
 Here $h_1^0$ and $h_2^0$ are
 the neutral components of $h_1$ and $h_2$ respectively,
 and for simplicity we assume that these v.e.v.'s are real.

The most useful upper bound is obtained
 by the (1,1) component of the neutral scalar matrix $M^2$
 in the basis $(\phi_1,\phi_2,\phi_3)$ where $\phi_2$ does not
have any vacuum expectation value:
\begin {equation}
\left( \begin{array}{c}
                \phi_1\\
                \phi_2\\
                \phi_3
                \end{array} \right)
=\left( \begin{array}{ccc}
                \cos{\beta}  & \sin {\beta}& 0\\
                -\sin {\beta}& \cos{\beta} & 0\\
                0            & 0           & 1
                \end{array} \right)
\left( \begin{array}{c}
                h_1\\
                h_2\\
                n
                \end{array} \right) ,
\end {equation}
 with $\tan\beta = {v_2/ v_1}$ .
The upper bound of the lightest neutral scalar $S_1$ is given by
\begin {equation}
m_{S_1}^2\leq M_{11}^2
 = m_Z^2 \cos^2{2\beta}+\lambda_1^2 v^2 \sin^2{2\beta}+\delta ,
\end {equation}
where $v=\sqrt{v_1^2+v_2^2}=174 $ GeV and
 $\delta$ represents the top and stop effects
\begin {equation}
\delta=
\frac{3}{4\pi^2}\frac{m_t^4}{v^2}
\ln{\frac{m_{stop}^2}{m_{t}^2}}_{\,\,\, .}
\end {equation}
The maximum value of $\lambda_1$
 is determined by requiring
 that none of the dimensionless coupling constants
 blow up below the GUT scale ($10^{16}$ GeV).
 In figure 1 we present the upper bound of eq.(5)
 in the parameter space of the top mass
 and $\tan{\beta}$.
 The corresponding figure for the MSSM is also shown.
 We can see
 that the maximum value of the lightest Higgs is
 about 135 GeV which is realized
 in the parameter region
 of a large $\tan{\beta}$ and large $m_t$
 or of a small $\tan{\beta}$ and small $m_t$.
 Note that the $\delta$ term in eq.(5) becomes large
 in the former case,
 on the other hand the $\lambda_1$ term in eq.(5)
 becomes more important in the latter case.
 This should be compared to the MSSM case
 where the maximal value is given only
 in the large top mass region.

As is explained before it is important to determine
 upper bounds on the heavier neutral Higgs masses
 because the lightest one may be dominated by the singlet.
 We can derive an upper bound on the second
lightest neutral mass in the following way.
 Denoting the second lightest and the heaviest
 neutral scalar Higgs by $S_2$ and $S_3$ respectively,
 the orthogonal diagonalization matrix
 $V$ is given by
\begin {equation}
\left( \begin{array}{c}
                \phi_1-v\\
                \phi_2\\
                \phi_3-x
                \end{array} \right)
= V
\left( \begin{array}{c}
                S_1\\
                S_2\\
                S_3
                \end{array} \right)_{\,\,\, .}
\label{eqn:basis}
\end {equation}
 It is a matter of a simple algebra to show
 that the second lightest
 eigenvalue satisfies the following inequality:
\begin {equation}
m_{S_2}^2\leq \frac{M_{11}^2-V_{11}^2 m_{S_1}^2}{1-V_{11}^2}_{\,\,\, ,}
\label{eqn:upbnds2}
\end {equation}
where $V_{11}$
is the (1,1) component of the diagonalizing matrix.
$V_{11}$ is directly related to the following quantity:
\begin {equation}
V_{11}^2=\frac {\sigma(e^+ e^- \rightarrow Z^0 S_1)}
          {\sigma(e^+ e^- \rightarrow Z^0 h_{SM})}
          \Bigl|_{m_{S_{1}}=m_{h_{SM}}\,\,\, ,}
\end {equation}
where $\sigma(e^+ e^- \rightarrow Z^0 h_{SM})$
 is the production cross section for the standard model Higgs
 and $ \sigma(e^+ e^- \rightarrow Z^0 S_1)$
is the corresponding production cross section
 for the lightest neutral Higgs boson
 in the present model.
 In figure 2 we show a contour plot of the upper bound on
 the second lightest Higgs mass
 in the space of the lightest neutral Higgs mass
 and $V_{11}^2$.
 In this figure we take the top mass to be 150 GeV
 and the stop mass to be 1 TeV
 and the resulting upper bound
 of the lightest Higgs mass is 132 GeV
 which is realized at $\tan\beta=1.9 $.
 We can see that
 the upper bound on the second lightest Higgs mass can be 10 \%
larger than $ M_{11}$ in the region where $V_{11}^2 \lsim 0.2$.
 This means that even
 when the lightest neutral Higgs is singlet-dominated,
 and $V_{11}^2$ is thereby reduced, we still can expect at least
 one doublet-like Higgs below, say,
 150 GeV$\,$\footnote{It is pointed out in ref.~\cite{Elliot} that
 the mass of the second lightest Higgs
 satisfies the upper bound of the lightest Higgs mass
 in the limit that the lightest Higgs becomes the pure singlet.
 This is because the mass matrix for the neutral scalar Higgses
 becomes block diagonal.}.
 On the other hand, $m_{S_{2}}$ is not constrained
 in the limit of $V_{11}\rightarrow 1$ .

When the second lightest Higgs as well as the first one
 has a reduced coupling to the $Z^0$ boson
 neither of them can be produced\footnote{
 In this case the {\it scalar-pseudoscalar}-$Z^0$ coupling may become
 significant, and the scalar-pseudoscalar pair production
 is possible if the pseudoscalar boson is light enough.}.
 Then we need an upper bound of the heaviest neutral Higgs boson mass.
  A simple generalization of eq.(8) gives us
 the following relation for the heaviest neutral Higgs mass:
\begin {equation}
m_{S_3}^2\leq \frac{M_{11}^2-(V_{11}^2+V_{12}^2) m_{S_1}^2}
{1-(V_{11}^2+V_{12}^2)}_{\,\,\, .}
\label{eqn:upbnds3}
\end {equation}
Then the contour plot for the upper bound of the heaviest neutral Higgs boson
mass is the same as the previous one provided
 that the x-axis is replaced by $V_{11}^2+V_{12}^2$.

In order to show usefulness of the above formulas
we consider the Higgs detection in a future
  $e^+ \, e^-$  linear collider with $\sqrt{s}\sim 300$ GeV.
 The neutral scalar Higgses can be produced
 by either $e^+ e^-\rightarrow Z^0S_i$ process
 or $e^+e^-\rightarrow A^0_iS_j$ process
 where $A_i^0$ is a pseudoscalar Higgs.
 Since the availability of the latter process depends on
 the mass of the pseudoscalar boson
 we only consider here the former process
 as a production mechanism of the neutral Higgs.
 Furthermore the decay modes of the neutral scalar Higgs
 are quite parameter-dependent.
 Although the standard model Higgs of the intermediate mass decays
 predominantly into a $b\overline{b}$ pair,
 it is known in the MSSM case that
 the main decay mode of the neutral Higgs may be
 an invisible neutralino pair depending on parameters\cite{Zerwas,Aleph}.
 A similar situation can be realized in the present model.
 Even in such a case the recoil mass distribution measurement
 in the process $e^+ e^-\rightarrow Z^0S_i$ is available to show
 an evidence for the neutral Higgs.
 Therefore, we assume that the neutral Higgs boson
 can be detected if the production cross section for
 $e^+e^-\rightarrow Z^0S_i$ mode is large enough.
The production cross section for $S_i$ is obtained from
that for the standard model Higgs of
the same mass by multiplying the $V_{1i}^2$ factor.
The standard model cross section is given by\cite{HHG}
\begin {equation}
\sigma_{SM}= \frac{\pi \alpha^2(1+(1-4\sin^2{\theta_W})^2)}
                  {192 s \sin^4{\theta_W}\cos^4{\theta_W}}
             \frac{(\lambda+12m_Z^2/s)\sqrt{\lambda}}
                  {(1-m_Z^2/s)^2} ,
\end {equation}
where $\lambda$ is
\begin {equation}
\lambda= \left[1-\frac{(m_Z+m_h)^2}{s}\right]
 \left[1-\frac{(m_Z-m_h)^2}{s}\right] ,
\end {equation}
and $m_h$ is the Higgs mass.

Let us now consider the detectability of the three neutral scalar Higgses.
We would like to know whether at least one of the three Higgses
can be discovered at an $e^+ \, e^-$  linear collider
with a given integrated luminosity.
 For a given set of $V_{11}^2,V_{12}^2$ and $m_{S_1}$,
 we can determine the upper bounds on
 $m_{S_2}$ and $m_{S_3}$ from eqs.(\ref{eqn:upbnds2}) and (\ref{eqn:upbnds3}),
 then we can calculate
 the lower bounds of the production cross section
 for the second and third Higgses
 as well as the production cross sections
 for the lightest one.
 By denoting the above three values of
the cross sections as $\sigma_{2min},\sigma_{3min},\sigma_1$ we then
define the following quantity:
\begin {equation}
\sigma(V_{11}^2,V_{12}^2)\equiv
 \min_{m_{S_1}}\{\max(\sigma_1,\sigma_{2min},\sigma_{3min})\}.
\end {equation}
The meaning of this value is that for a given point of the parameter
space ($V_{11}^2, V_{12}^2$)
 at least one of the three Higgses has always
a production cross section larger than
 $\sigma(V_{11}^2,V_{12}^2)$.
 In figure 3 we show
 the value for $\sigma(V_{11}^2,V_{12}^2)$
 in the space of $V_{11}^2$ and $ V_{12}^2$ for
$\sqrt {s}=300$ GeV, $m_t$ = 150 GeV and $m_{stop}$ = 1 TeV .
 The minimum of the cross section is 0.046 pb.
This corresponds to about 30 events of  $ e^+e^-\rightarrow Z^0S_i$
followed by $Z^0\rightarrow e^+e^-\, ,\, \mu^+\mu^-$
in integrated luminosity of 10 fb$^{-1}$ which will be attained
 at JLC-I after the operation of one third of a year\footnote{
 A more realistic estimation taking account of detection efficiency
 will increase the necessary integrated luminosity to
 about 30~fb$^{-1}$\cite{JLC,MIYAMOTO}}.
 Since the production cross section
 of the standard model Higgs of the same mass as
 the lightest neutral Higgs' upper bound (132 GeV) is 0.17 pb,
 the production cross section
 is reduced by a factor 4 at the worst case and
 therefore four times as large integrated luminosity is required
 to cover the whole parameter space.
 In the above analysis we made essential use of
 eqs.(\ref{eqn:upbnds2}) and (\ref{eqn:upbnds3}) where
 the upper bounds of the second and the third Higgs masses are given
 in term of $V_{11}^2$ and $V_{11}^2+V_{12}^2\,$.
 Suppose that neutral Higgses are not found
 below the lightest upper bound for an enough integrated luminosity.
 This means that the lightest one has a reduced coupling to the $Z^0$
 boson, therefore we get an upper bound on $V_{11}^2$.
 Then, from the figure 2,
 we can determine an upper bound on the second lightest Higgs mass.
 If the second Higgs is also not found up to its upper bound,
 we can use a similar argument and
 determine upper bounds on $V_{12}^2$ and
 the third Higgs mass.
 Since we now know the upper bound of the third Higgs mass and
 the lower bound of $V_{13}^2$ from
 the relation $V_{13}^2=1-V_{11}^2-V_{12}^2$, we can determine
 whether the third Higgs
 can be discovered for the given integrated luminosity.

In figure 4 we show
 the minimum value of $\sigma(V_{11}^2,V_{12}^2)$
 in the ($V_{11}^2, V_{12}^2$) space,
 denoted by $\sigma_{min}$,
 as a function of $\sqrt{s}$ for various
top masses and a stop mass of 1 TeV.
 We can see that we always have
a neutral scalar Higgs with a production cross section
larger than $0.04\,$ pb around $\sqrt{s}= 300$ GeV.
 This is encouraging
 for a future $e^+e^-$ linear collider like JLC-I
 since independently of parameters in the model
 at least one of the neutral scalar Higgses is detectable at the level of
 one year of operation and
 this can be a definitive test of this model.

 Finally, we make comments on several possible extensions.
 The value of $M_{11}$ may be changed
 by the left-right stop mixing
 and the $\lambda_{1}, \lambda_{2}$ loop effects.
 However once the value of $M_{11}$ is specified
 the argument after eq.(\ref{eqn:basis})
 does not change. Since these effects do not modify
 $M_{11}$ significantly,
 we expect a similar conclusion
 on the production cross section of the neutral scalar Higgses.
 We may also consider some other extensions
 of the Higgs sector.
 When the v.e.v's become complex or
 more additional singlets are included,
 the dimension of the relevant mass matrix increases.
 A straightforward generalization
 of the above analysis
 is possible in such extensions.

The authors would like to thank
 K.~Hikasa, T.~Kawagoe, A.~Miyamoto and T.~Yanagida
 for useful discussions and comments.
 This work is supported in part by the Grant-in-aid for
Scientific Research from the Ministry of Education,
 Science and Culture of Japan.

\newpage
\noindent{\large {\bf Figure Captions}}\\
Fig.1: (a) The contours of the lightest Higgs mass upper bound
 in the space of the top mass and $\tan\beta$
 for the model with a gauge singlet Higgs.
 The {\it blow up region\/} is excluded
 since the top Yukawa coupling blows up below the GUT scale
 even if $\lambda_1=\lambda_2=0$.
 (b)~The corresponding contours for the MSSM. The blow up region is not
 shown in this figure.
\\
Fig.2: The contours of the second light Higgs mass upper bound
 for $m_t=150$ GeV and $m_{stop}=1$ TeV.
\\
Fig.3: The contours of $\sigma(V_{11}^2,V_{12}^2)$
 in the $V_{11}^2$-$V_{12}^2$ plane.
 We take $\sqrt s= 300$ GeV, $m_t=150$ GeV and
 $m_{stop}=1$ TeV.
\\
Fig.4: The minimum cross section $\sigma_{min}$
 for the top masses, $m_t=$120, 150 and 180 GeV,
 and $m_{stop}=1$ TeV.
\\
\end{document}